\documentclass[%
 reprint,
 amsmath,amssymb,
 aps,
]{revtex4-1}

\usepackage{graphicx}% Include figure files
\usepackage{dcolumn}% Align table columns on decimal point
\usepackage{bm}% bold math
\begin{document}

\preprint{APS/123-QED}

\title{Electronically Steerable Ultrasound-Driven Air Stream}% Force line breaks with \\

\author{Keisuke Hasegawa$^*$, Liwei Qiu$^*$, Akihito Noda$^\dagger$, Seki Inoue$^\ddagger$, Hiroyuki Shinoda$^*$}
%\author{Akihito Noda}%
\affiliation{%
$^*$Graduate School of Frontier Sciences, The University of Tokyo\\
$^\dagger$Department of Mechatronics, Nanzan University\\
$^\ddagger$Graduate School of Information Science and Technology, The University of Tokyo
}%


\date{\today}% It is always \today, today,

\begin{abstract}
We have created a stretching air flow by generating a Bessel beam of ultrasound with an active phased array of acoustic transducers in free space.
The generated Bessel beam is electronically steerable in terms of its position and direction of propagation.
The fastest spot in the flow could be located apart from the sound source itself.
The flow is formed immediately after the sound field is generated, which can be electronically accomplished with the phased array that we fabricated.
We formulate the basic technique for realizing such air-borne flows and experimentally describe their physical properties.
Our technique would be capable of controlling the air flow in free-space with an improved temporal and spatial resolution compared with those conventionally considered.

\begin{description}
\item[Keyword]
Nonlinear Acoustics, Ultrasound, Fluid Dynamics
% \item[PACS numbers]
% May be entered using the \verb+\pacs{#1}+ command.
% You may use the \texttt{description} environment to structure your abstract;
% use the optional argument of the \verb+\item+ command to give the category of each item. 
\end{description}
\end{abstract}

\maketitle

%\tableofcontents

Recently mid-air nonlinear acoustics in free space has drawn considerable interests of researchers among various disciplines, from the perspective of its scientifically intriguing properties and potential practical applications.
They include mid-air particle trapping\cite{ono}, \cite{hoshi1}, contactless vibro-tactile stimulation\cite{hoshi2}, \cite{hasegawa}, remote mechanical sensing\cite{fujiwara} and so forth.

Those applications are mostly based on the acoustic radiation pressure, which is one of the representative nonlinear acoustic phenomena.
It is a static pressure in macroscopic temporal scale that can be observed on the medium boundary where the sound propagation is blocked\cite{awatani}.
On the other hand, there is still another significant nonlinear-effect in acoustics called acoustic streaming\cite{hamilton}.
It refers to a flow of mass in the air that arises with the intense sound waves.
In case with sinusoidal sources, the driving force of the flow is seen parallel to sound propagation \cite{kamakura}.
This phenomenon has been known and investigated through decades, while its applications seem not to have been sought as much as that related to the acoustic radiation pressure.

The main idea of the research is that environmental air flows in free-space can be controlled by steered sound propagations.
One of the interesting and useful properties expected in this sound-driven air flow is a narrow flow with its depth taking its fastest can be set apart from the sound source.
It is antithetical to the normal jet-driven flows with its streaming velocity tapering off.
What makes it possible is the fact that acoustic energy can be spatially localized far from the source with such techniques as acoustic lens \cite{acousticlens} or phase-controlled sources \cite{hoshi1}.
Since the spatial resolution of the wave field intrinsically depends on the wavelength, use of sources with higher frequency would yield narrower flows.
In addition, with an appropriately controlled acoustic source, the velocity and direction of the flow can be changed with the latency of speed of sound.
Owing to the nature of sound, when the sound propagation is reflected on the wall, a flow that comes from the wall can even be created.
Thus spatial and temporal resolution of the air flow in free space with an improved spatial flexibility is what we expect with our technique proposed in the following of the paper.

In order to generate a stretching flow, we exploit an ultrasound Bessel beam.
A Bessel beam is originally discussed in optical disciplines\cite{bboriginal}.
Its best-known property is non-diffracting propagation, where most of the energy emitted from the source is centralized around the propagation axis through a considerably long distance.
Therefore it is expected that a straight acoustic Bessel beam concurrently accompanies the straight flow inside it.

Converting the incoming planer wave with an optical axicon lens into Bessel beam would be the most prevalent method for obtaining optical Bessel beams.
There have been other methods proposed such as use of micromirrors\cite{micromirror}, meta-surface\cite{localmodes}, surface plasmon\cite{monnai} or antenna arrays\cite{carrays}.

Beside those optic/electromagnetic domains, Bessel beams in acoustics have also been created with a variety of principles. They include specifically designed speaker\cite{bbeam}, acoustic axicon-lens\cite{axicon}, grated concentric source arrays\cite{grated} and the use of acoustic meta-surface\cite{meta}.

\begin{figure}[t]
\includegraphics[width=84mm]{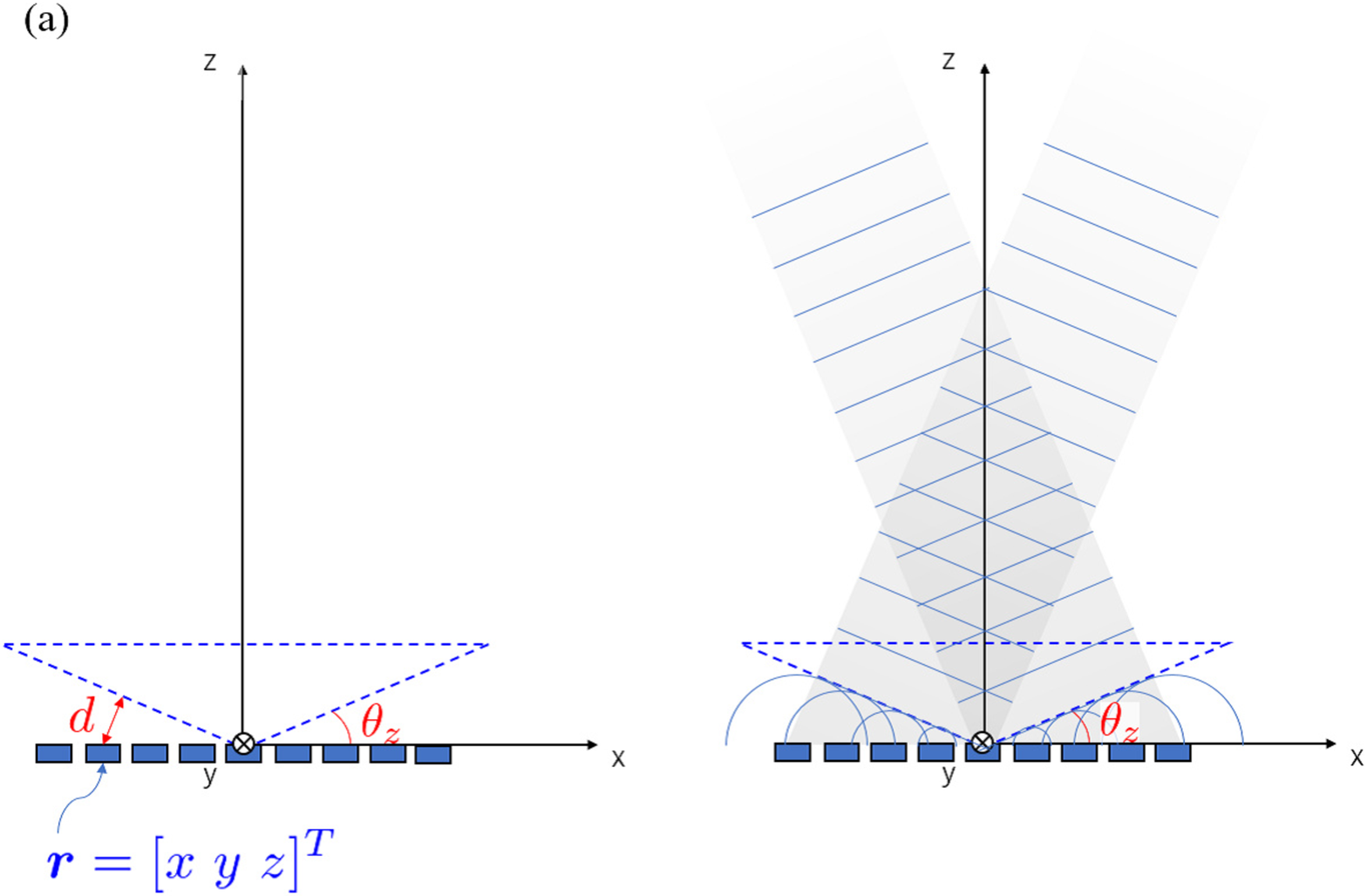}\\
\includegraphics[width=84mm]{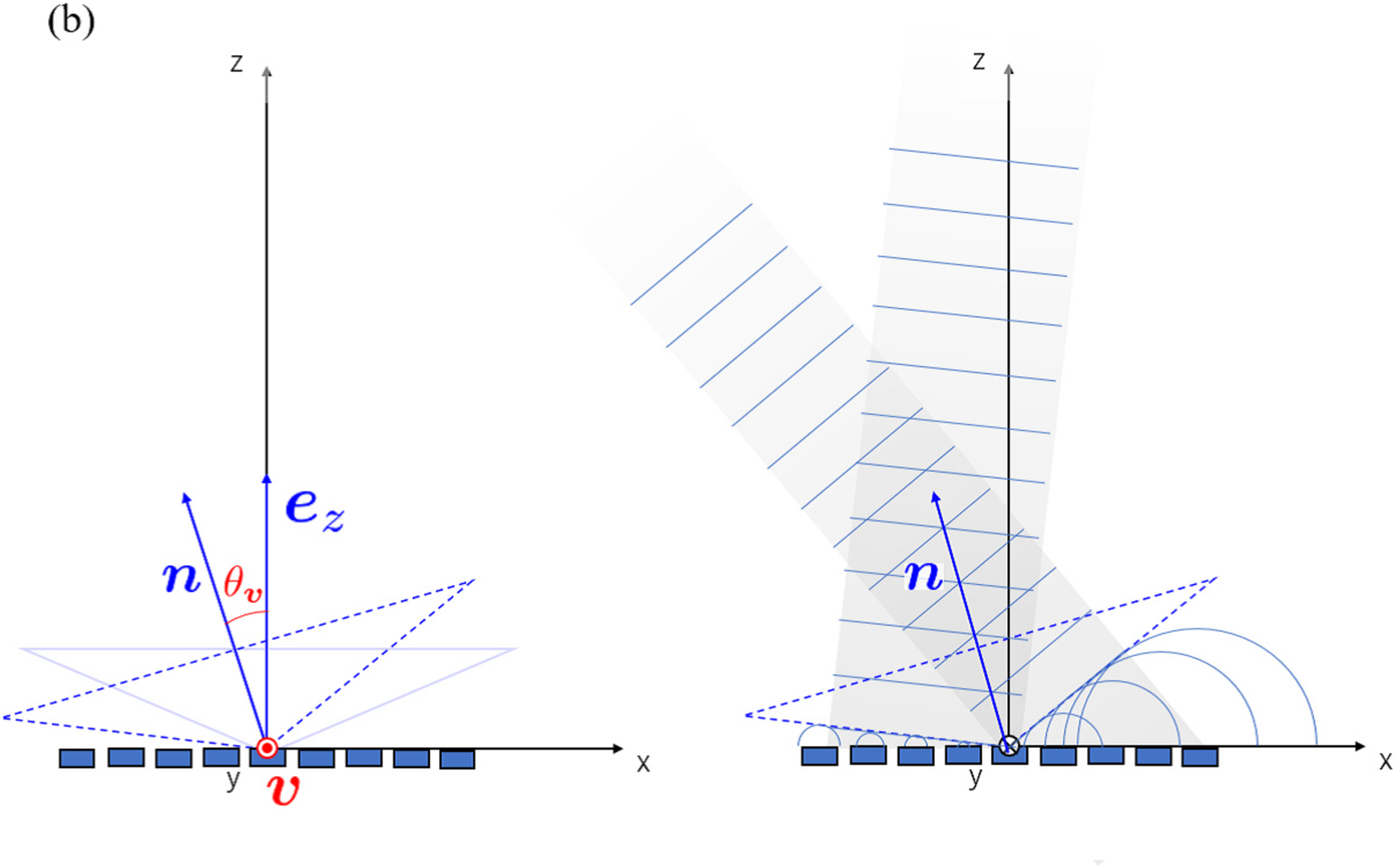}% Here is how to import EPS art
\caption{Virtual cone-shaped sound source and collective wavefront emitted from arrayed multiple point sources when the resulting Bessel beam is: (a) perpendicular to the array and (b) tilted.}
\end{figure}

Most of those technique has a common issue that the generated Bessel beams have a spatial profile that is fixed or hardly tunable.
The problem is derived from the fact that once the geometry of the sound source is fixed, spatial property of sound propagation is essentially determined.
This would be problematic when one wants a beam whose position and direction is steerable, which limits the range of applications.

One solution for the problem would be the phased-array technique, which integrates multiple coherent sources each of whose output phase delay is electronically controllable.
This technique has widely been used in the realm of ultrasound engineering\cite{parray}.
With a proper set of phase-delayed emissions from multiple sources, the wavefront of generated ultrasound can be controlled.
It means that we can design the orientation of ultrasound propagation electronically, which would realize steerable sound-driven air streaming.

We describe a method for generating tilted acoustic Bessel beam out of ultrasound phased array technique.
With an analogy to the axicon lens, sources with phase shifts proportional to distance from the center of the source creates a Bessel beam.
This is intuitively understood as those phase shift having a role as a tilted irradiation face of an axicon lens.
Based on the principle of Huygens, multiple point wave sources that create a conical wave front can be considered as a virtual acoustic cone-shaped source resulting in acoustic Bessel beam that propagates toward the central axis of the cone.
Therefore each source located in $\bm{r} = \left[ x \ y \ z\right]^T$ should have the phase shift proportional to the distance from the virtual source cone (here $^T$ means the transpose of a vector).
Suppose that the $z$ axis is parallel to the axis of the cone, $\theta_z$ is the angle between the side of the corn and the $xy$-plane.
Geometrically, the difference between the corn and the source $d$ is given as
\begin{equation}
d = \sin\theta_z\sqrt{x^2 + y^2} - \cos\theta_z z.
\end{equation}
Note that the value of $d$ here is 'signed'.
It takes a negative value when $\bm{r}$ is 'inside' the cone, corresponding to giving negative phase shifts where the wavefront of the cone arises behind ultrasound sources.

Decreasing $\theta_z$ to zero means that the beam reaches farther.
At the same time, the generated wave front as a whole becomes more similar to a plane wave, resulting in less concentrated beam around the $z$-axis.
Therefore it can be said that there is a trade-off between the beam concentration and its reaching distance.

A case with a tilted Bessel beam is discussed below.
Let $\bm{n} = \left[ n_x \ n_y \ n_z\right]^T$ be a unit vector which indicates the direction of the propagation that holds $n_x^2 + n_y^2 + n_z^2 = 1$.
By tilting the cone axis $\bm{e}_z = \left[ 0 \ 0 \ 1\right]^T$ toward $\bm{n}$, a Bessel beam orienting $\bm{n}$ can be obtained.
The corresponding rotational operation can be expressed by the rotation axis $\bm{v}$ with its angle $\theta_{\bm{v}}$.
These are given as 
\begin{eqnarray}
\bm{v} & = & \frac{\bm{e}_z \times \bm{n}}{|\bm{e}_z \times \bm{n}|}, \\
\theta_{\bm{v}} & = & \sin^{-1}\left(|\bm{e}_z \times \bm{n}|\right).
\end{eqnarray} 

In order to realize the tilted cone source, the difference calculation should be done with the transducer position that are rotated in line with the rotational operation described above.
According to the Rodrigues's rotational rule, the rotated transducer position $\bm{R} = \left[ X \ Y \ Z\right]^T$ by a rotational axis $\bm{v}$ with an rotational angle $- \theta_{\bm{v}}$ is obtained as

\begin{equation}
\bm{R} = \cos\theta_{\bm{v}}\bm{r} - \sin\theta_{\bm{v}}(\bm{v}\times\bm{r}) + (1 - \cos\theta_{\bm{v}})(\bm{v}^T\bm{r})\bm{v}.
\end{equation}

Therefore the proper phase delay $\phi$ that are imposed on the transducer is given as
\begin{eqnarray}
\phi & = & k\left(\sin\theta_z\sqrt{X^2 + Y^2} - \cos\theta_z Z\right),
\end{eqnarray}
where $k$ is the wavenumber of ultrasound.

In Fig. 1, a schematic explanation of the generated wavefronts with an array of sound sources is shown.
%The figure indicates that the sound field aiming to generate a tilted Bessel beam is expected to asymmetrical around the propagation axis.
%It is true in fact and the actual flow direction is biased as $\theta_{\bm{v}}$ increases. 

By substituting $x - x_0$ and $y - y_0$ for $x$ and $y$ respectively, one obtains a propagation axis rooted in $(x_0, y_0)$.
Although this might be evident, it is important to state that the beam does not have to be fixed at the center of the array from the perspective of application.

We measured the spatial distribution of an actual air flow that arises from ultrasound Bessel beam that we generated.
Figure 2 shows the experimental setup. 
We used multi-unit system of ultrasound phased array for generating ultrasound Bessel beam.
Each unit contained 249 ultrasound transducers.
The array was placed apart from the wall of the room in order to alleviate the effect of air circulation and to avoid generating standing waves.
The resonant frequency of the transducers was 40~kHz.
The maximum power consumption for each unit was approximately 50~W.
In the experiment we drove 4 units (Fig. 2(b)), resulting power consumption by 200~W at most.
The aperture constructed by the units was a rectangle of 373.8~mm(x-axis) x 292.7~mm(y-axis).
During all of the following measurements, we generated Bessel beams with the value of $\theta_z$ set to 18$^\circ$.
The streaming was observed immediately after the Bessel beam is emitted.

For the measurement of air flow velocity, we used an anemometer (KANOMAX CLIMOMASTER 6501-C0) attached on the hand of a robot (FANUC M710-iC 20L) scanning in two-dimensional planes.
The probe of the anemometer was KANOMAX 6533-21, which has no directivity around azimuthal angles.
It should be clearly noted that in the measurement we only captured the absolute value of flow velocity, not including its direction.
In the preliminary measurement the turbulence in the generated streaming was found to temporally range in several seconds.
Therefore we set the sensor output to be a temporal moving averages of the measurement during the most recent 10 second, which is converted into DC voltage.
Note that this fluctuation was not avoidable even averaging the output of the anemometer for one minute.
Our measurement consequently contains errors that are not too small to be negligible.

We performed the scanning with three set of measurement regions in order to evaluate the degree of spatial concentration in stream velocity for a non-tilted beam.
They are: (a) $xz$-plane where -200~mm~$\leq x \leq$~200~mm (every 20~mm), 100~mm $\leq x \leq$~1500~mm (every 100~mm),(b) $xz$-plane where 200~mm~$\leq z \leq$~600~mm (every 20~mm), -50~mm~$\leq x \leq$~50~mm (every 5~mm),  and (c) a plane where -20~mm~$\leq x \leq$~20~mm (every 2~mm), -20~mm$\leq y \leq$~20~mm (every 2~mm), $z = 400$~mm.
We also generated a tilted Bessel beam with the value of $\theta_{\bm{v}} = 20^\circ$ and $\bm{v} = [0 \ -1 \ 0]^T$, where the resulting wind velocity is scanned within the $xy$-plane where 200~mm~$\leq z \leq$~600~mm (every 20~mm), -200~mm~$\leq x \leq$~200~mm (every 20~mm).

\begin{figure}[t]
\includegraphics[width=80mm]{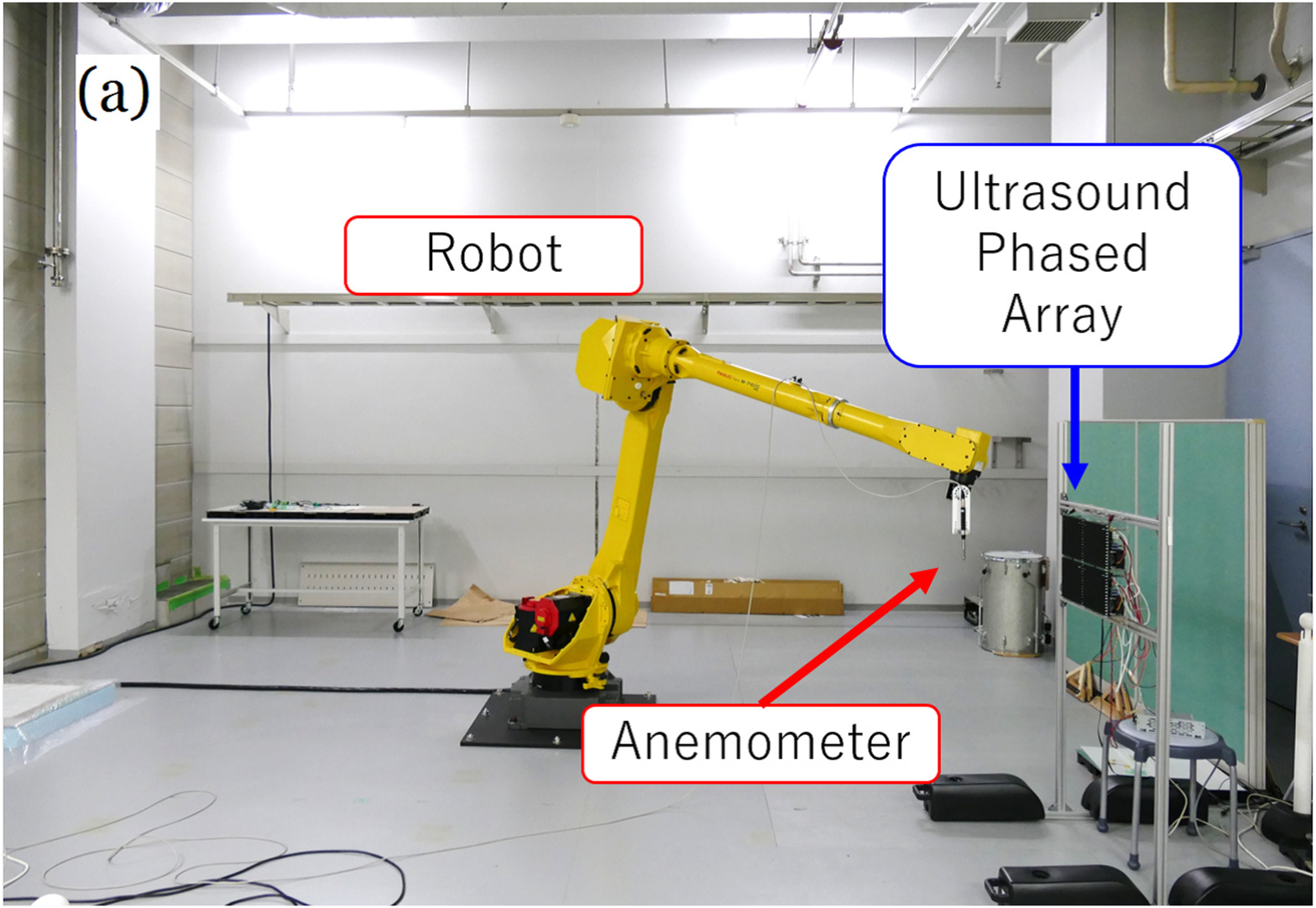}
\includegraphics[width=80mm]{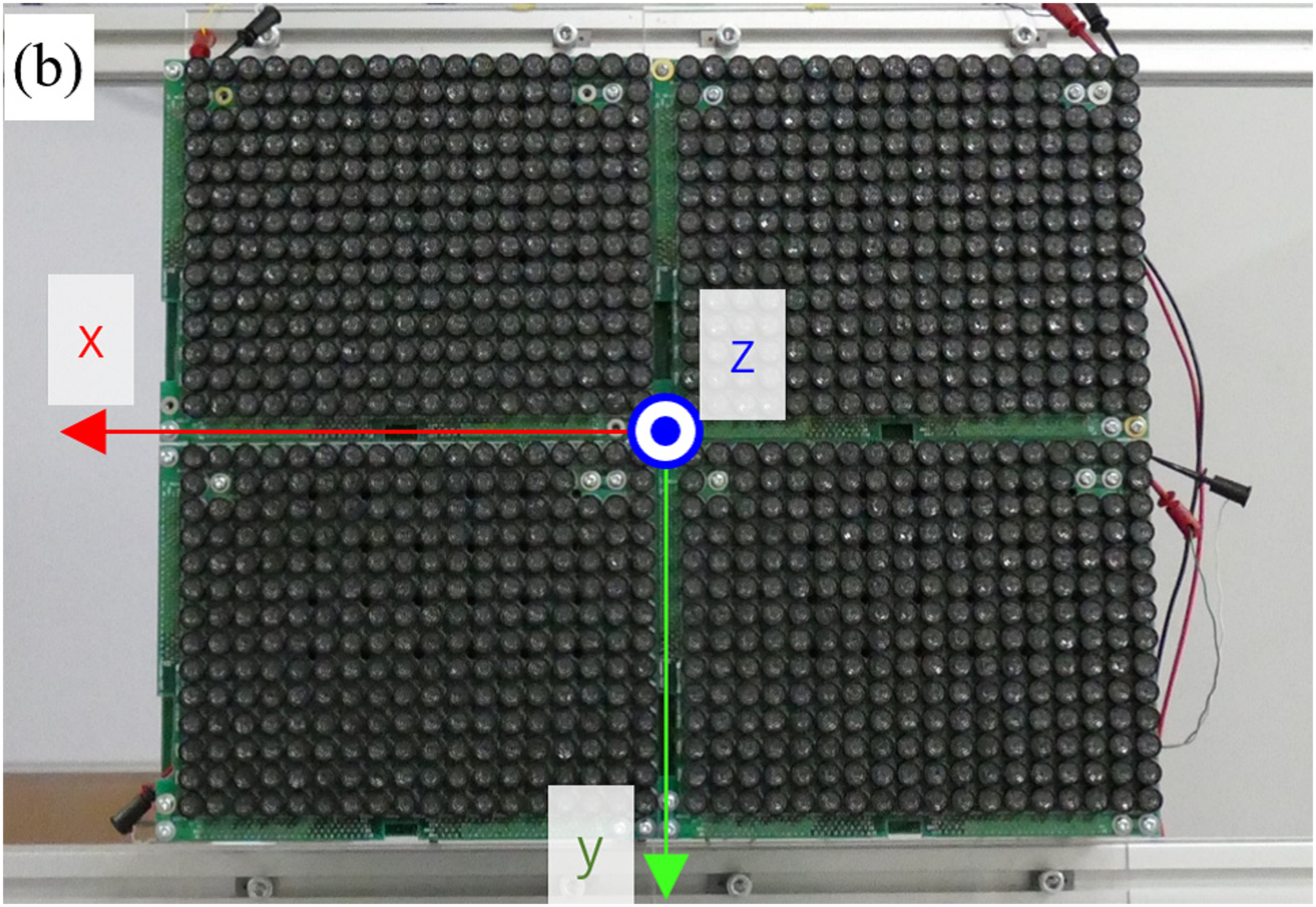}
\caption{(a): Measurement setup. (b): Ultrasound transducer array composed of four device units. The defined coordinates in the experiment is also depicted.}
\end{figure}

Figure 3 shows the measured wind velocity with the anemometer.
It is seen that the measured velocity takes its highest around $z = 400$~mm for the upright flow.
Thus the mid-air acceleration of the ultrasound-driven flow, which is never seen in ordinary jet-driven flows has been demonstrated.
In relation to the simulation result of acoustic pressure, spreading of the generated flow seems to be suppressed within the range of the acoustic beam.
On the other hand, spreading becomes obvious from where the flow reaches the end of the acoustic beam. 
This would be because of the beam-shaped driving force inside the streaming that stretches along the propagating axis.
It should be noted that at the most concentrating depth (z = 400~mm), most of the velocity of the flow is confined in the circle with its several millimeter of radius.
A flow concentrated to this degree from the source can be said to be a unique property of the ultrasound-driven streaming.

\begin{figure*}
\includegraphics[height=100mm]{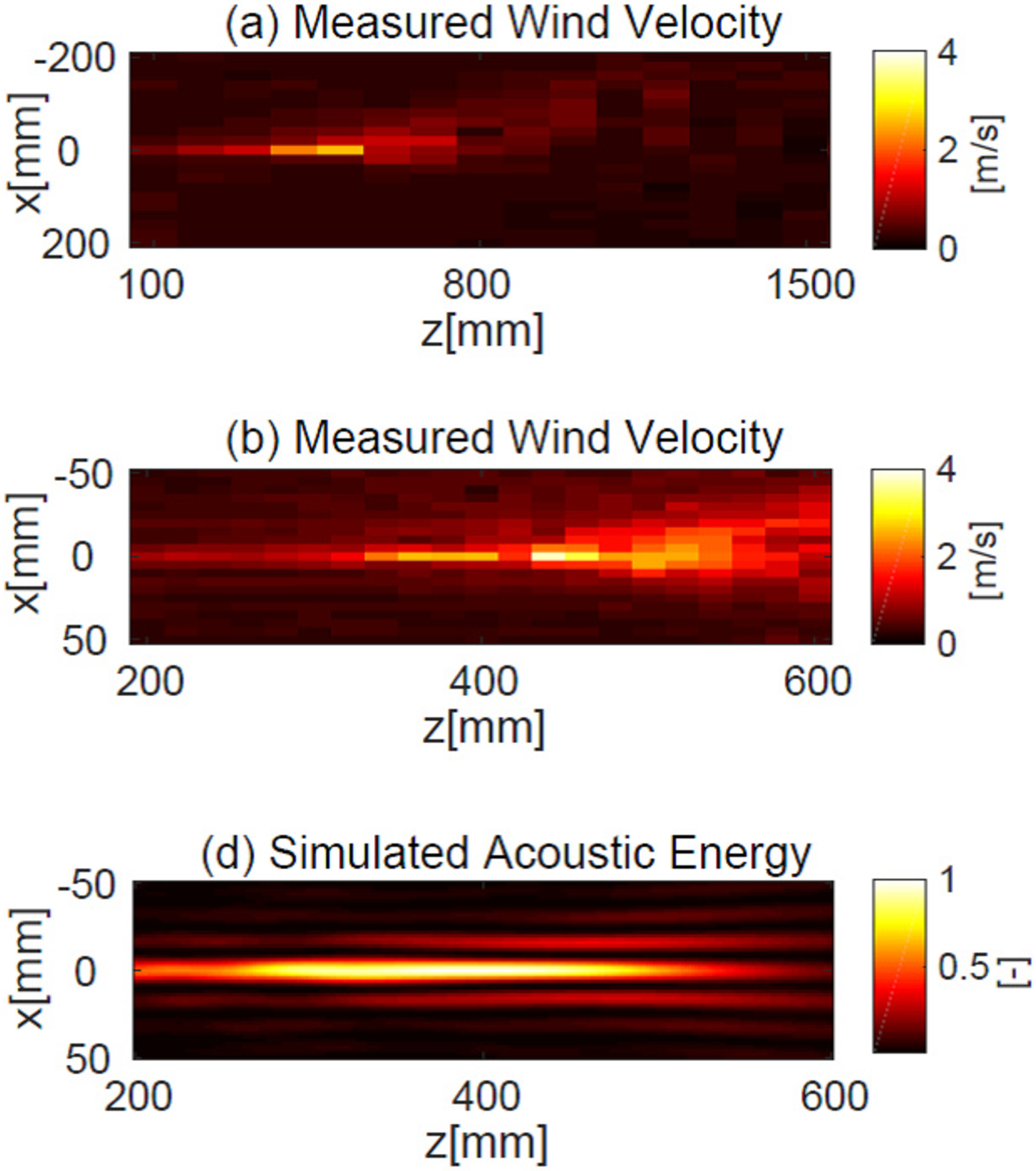}
\includegraphics[height=100mm]{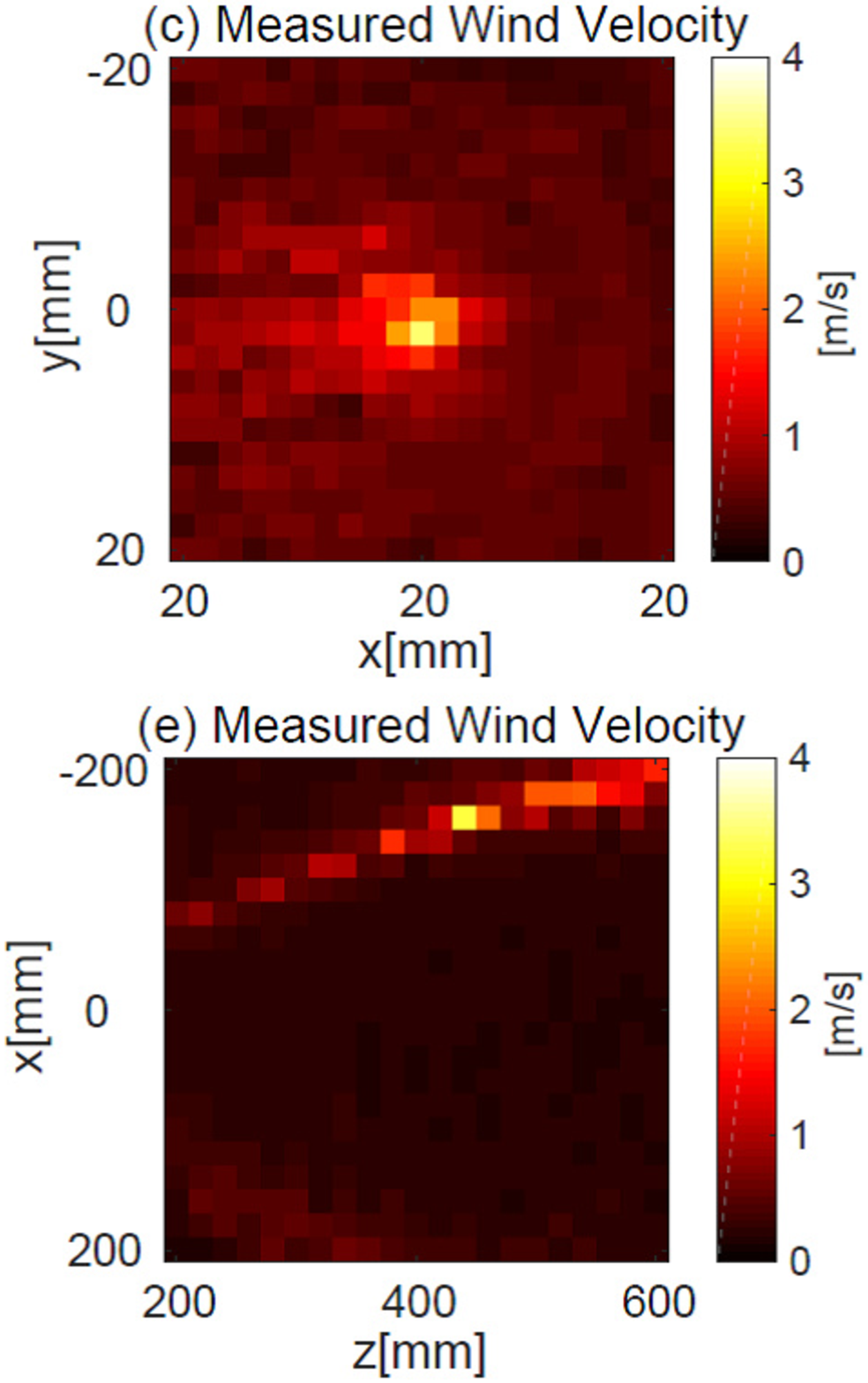}
\caption{(a), (b), (c):Measured spatial distribution of wind velocity in the case of a z-axis parallel beam. (d): Corresponding acoustic energy distribution estimated by a numerical simulation. (e): Measurement results with the same procedure in the case of a tilted beam ($\bm{v} = [0 \ -1 \ 0]^T, \theta_{\bm{v}} = 20^\circ$). $y = 0$~mm for (a),(b),(d),(e) and $z = 400$~mm for (c).}
\end{figure*}

Figure 3 (e) shows that a tilted flow is also successfully generated with the phase shifting technique described above.
Figure 4 shows the pictures of fog directed toward the beam propagation and numerical simulations indicating pressure of acoustic Bessel beam with the value of $\theta_{\bm{v}} = 0, 10^\circ, 20^\circ, 30^\circ$ where $\bm{v} = [0 \ -1 \ 0]^T$.
%It is seen that the less concentrated stream is sustained the more the Bessel beam is tilted, as exemplified in the case of $\theta_{\bm{v}} = 30^\circ$.
The simulations indicate the existence of ghost beams with different propagation should be seen.
In fact, a careful look on Fig.3(e) would find another weak flow toward different orientation that looks similar to Fig.4(h). 
Those 'grating lobes' is a common phenomenon created by a set of multiple wave sources.
The closer each source is arranged with one another, the more distant the intervals between each lobe become.
The simulations suggest that the resulting wavefront is perpendicular to the beam orientation.
Hence it is expected that the streaming is driven toward the stretching direction of the beam.
%As stated above, the sound field with tilted beam is expected to be asymmetrical around propagation axis.
%Numerical simulation indicates that the wavefront inside the beam is not perpendicular to the beam direction when the tilt increases.

\begin{figure*}
\includegraphics[width=33mm]{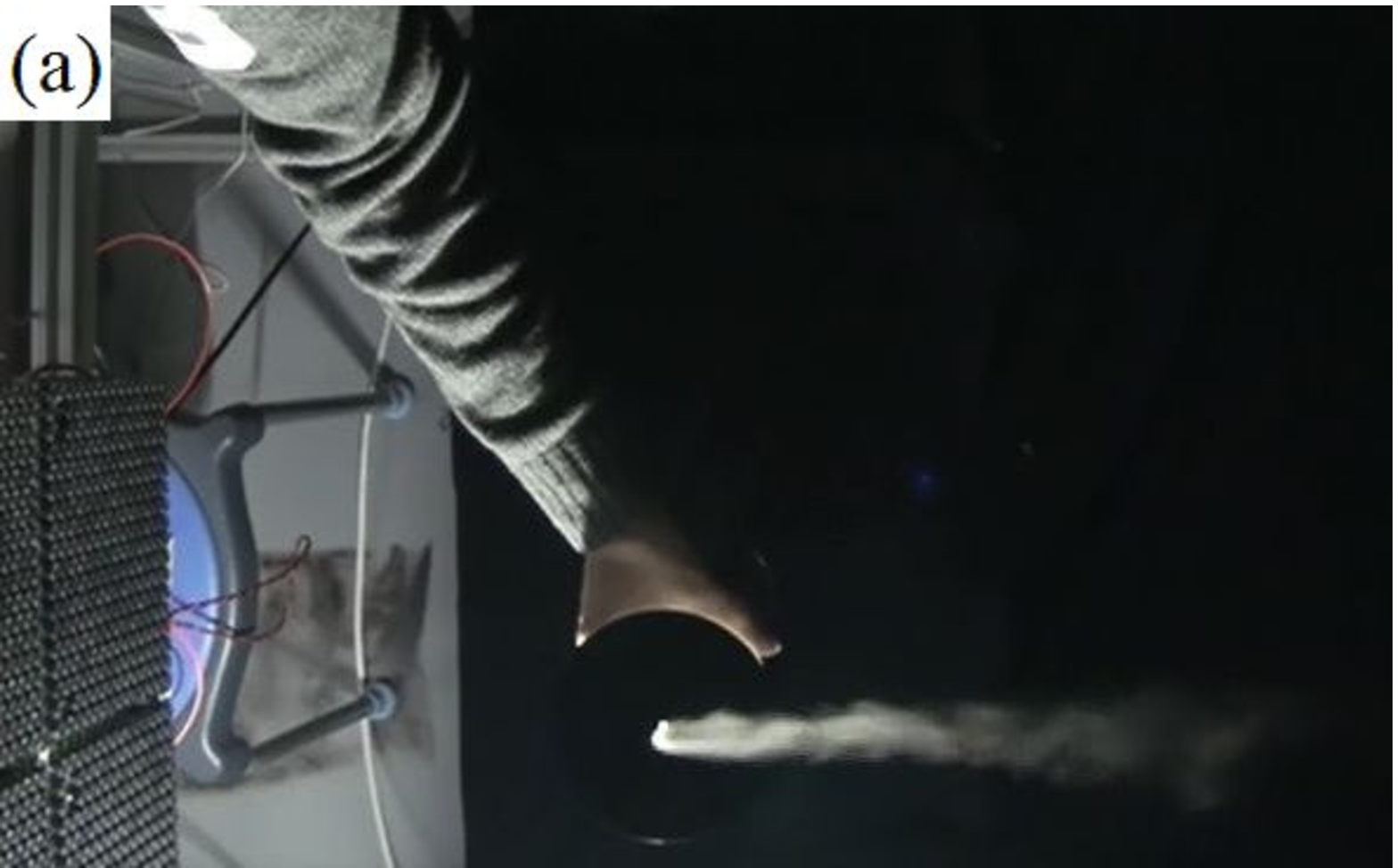}
\includegraphics[width=33mm]{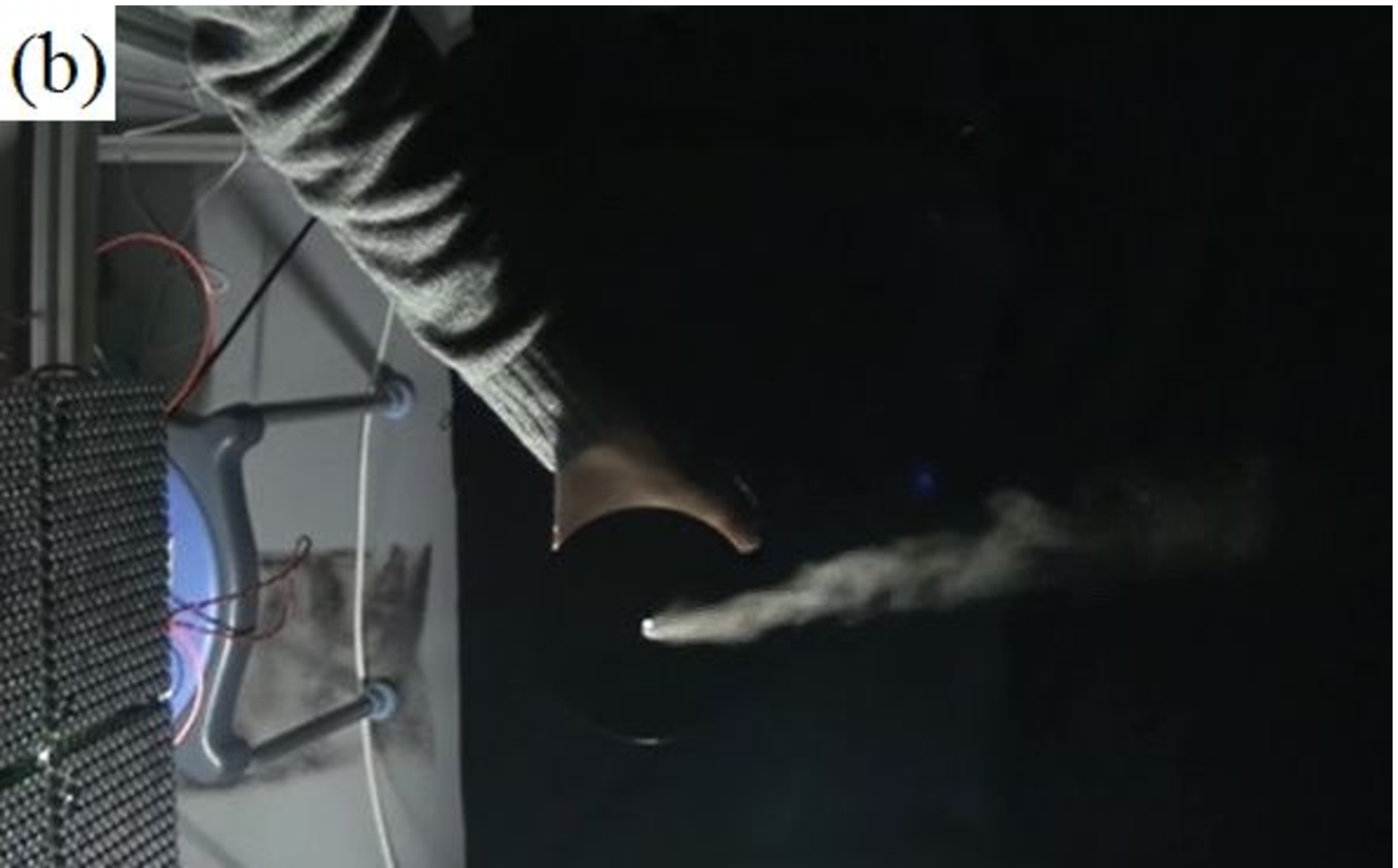}
\includegraphics[width=33mm]{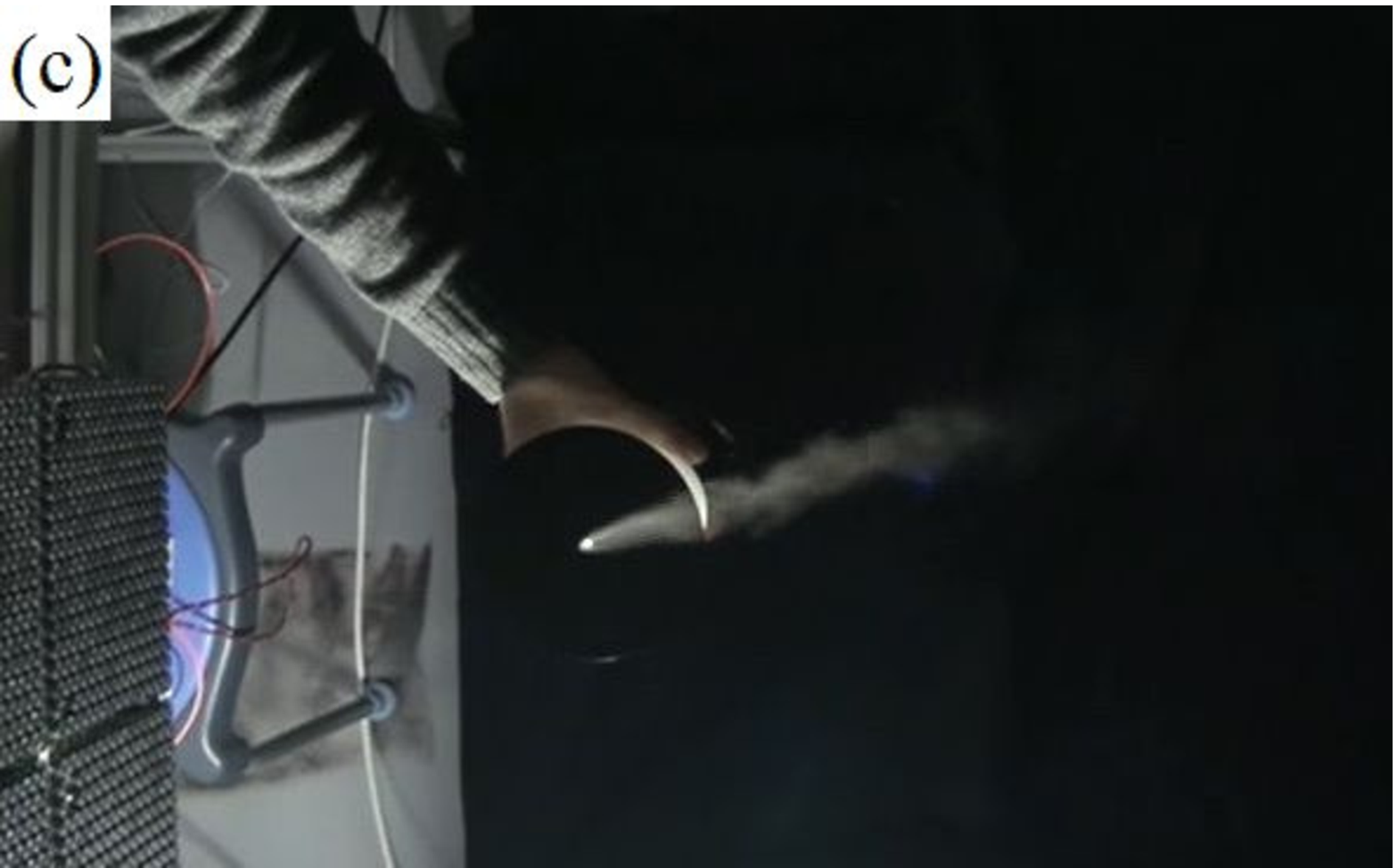}
\includegraphics[width=33mm]{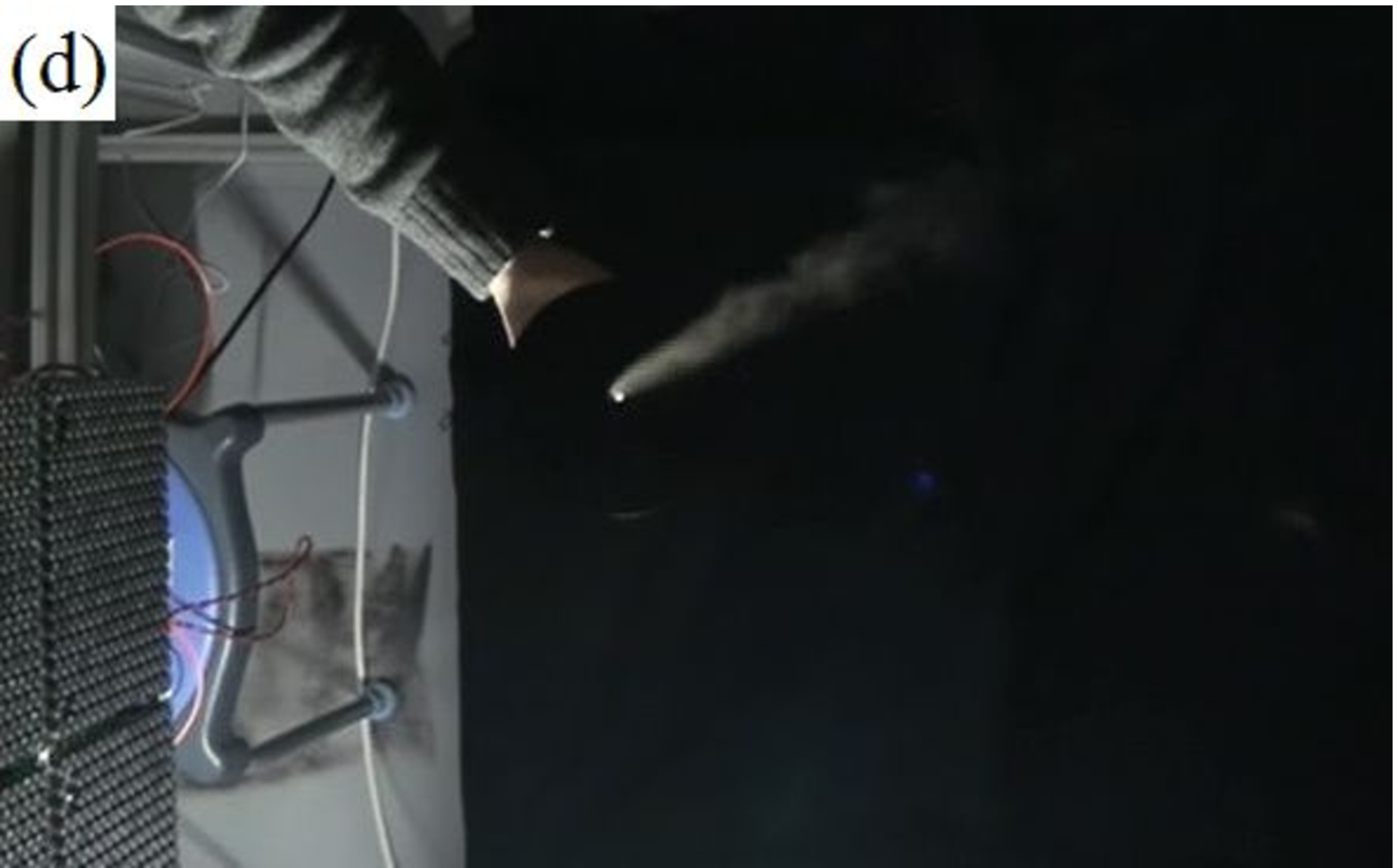}
\includegraphics[width=33mm]{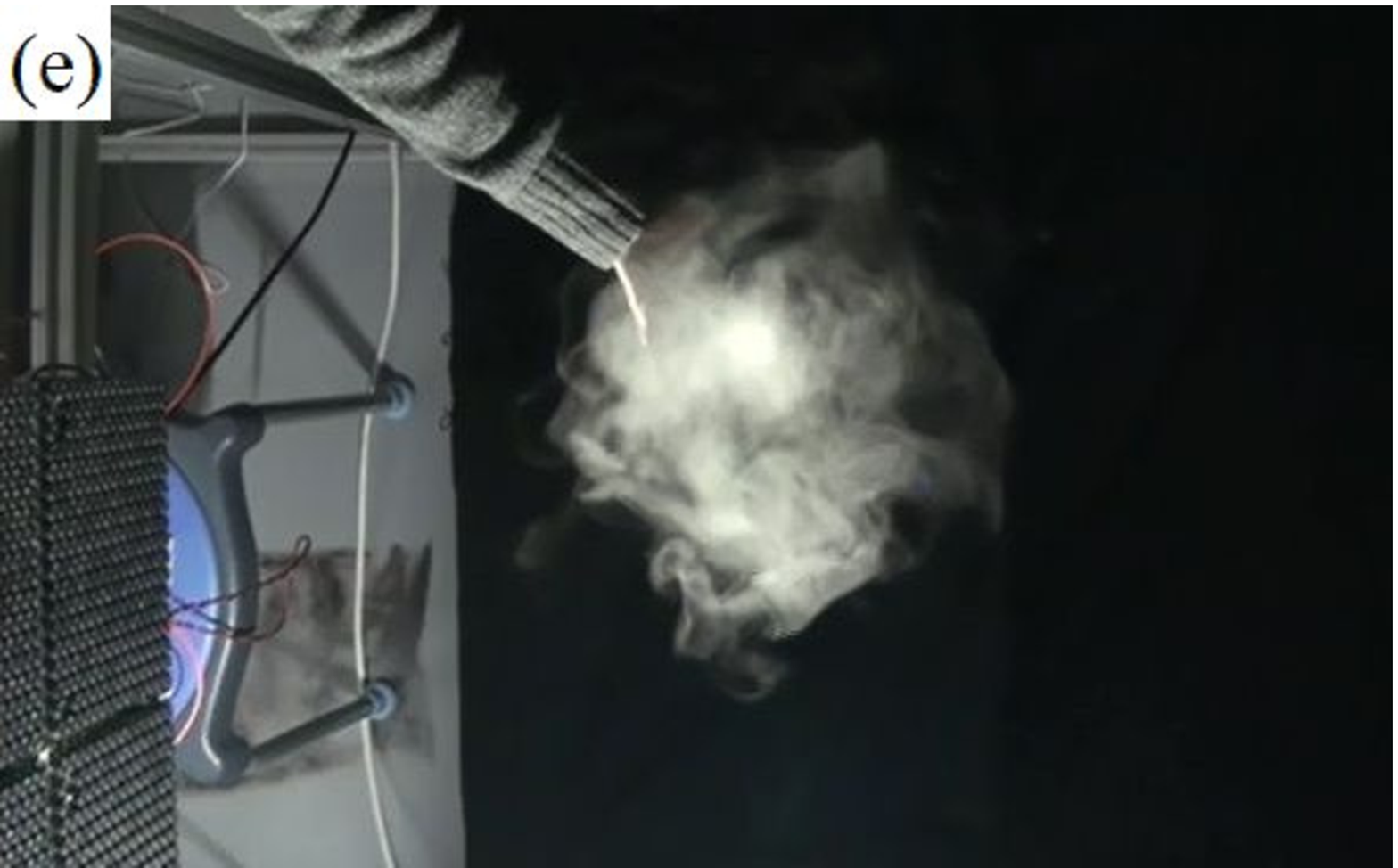}
\\
\includegraphics[width=33mm]{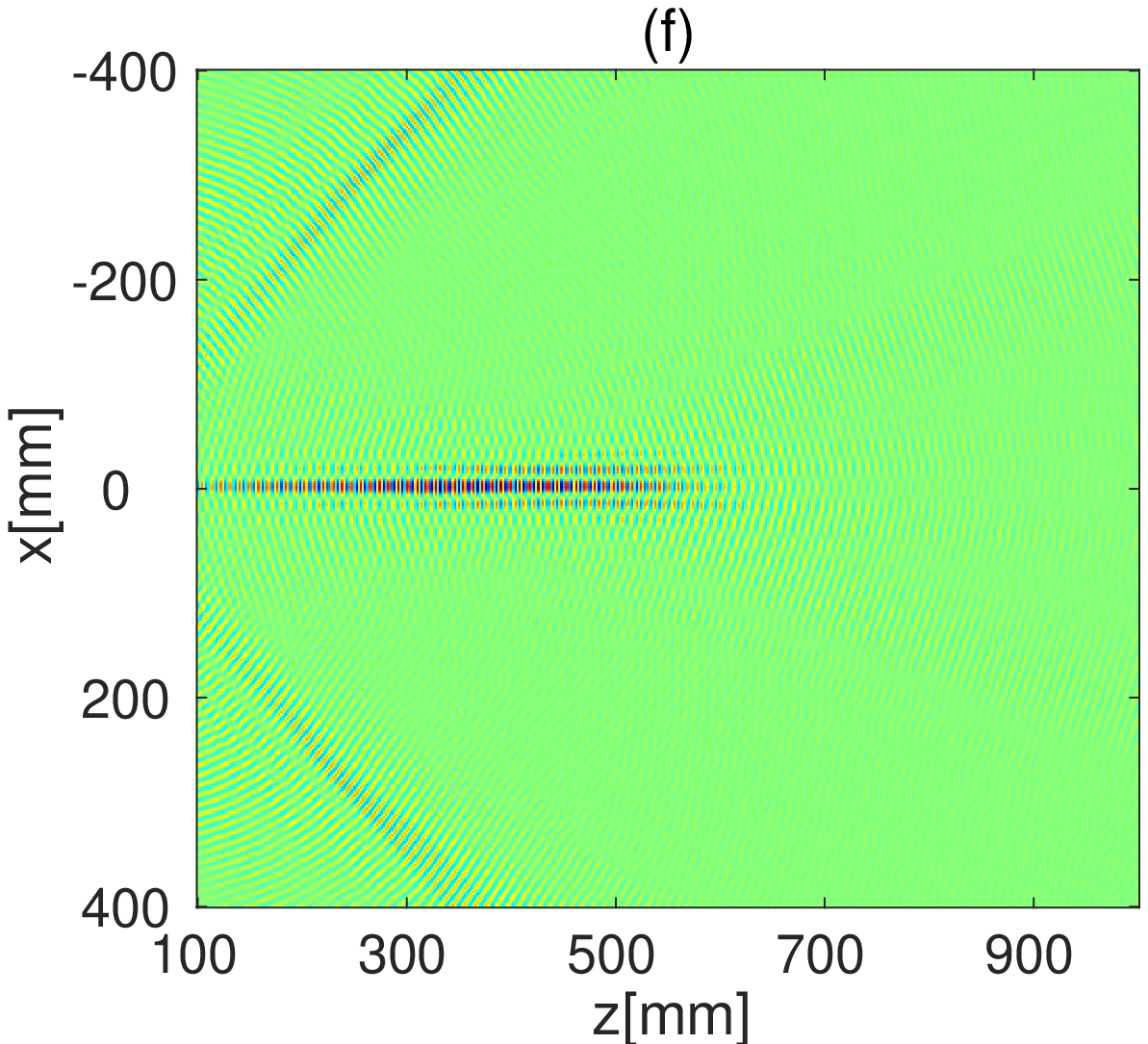}
\includegraphics[width=33mm]{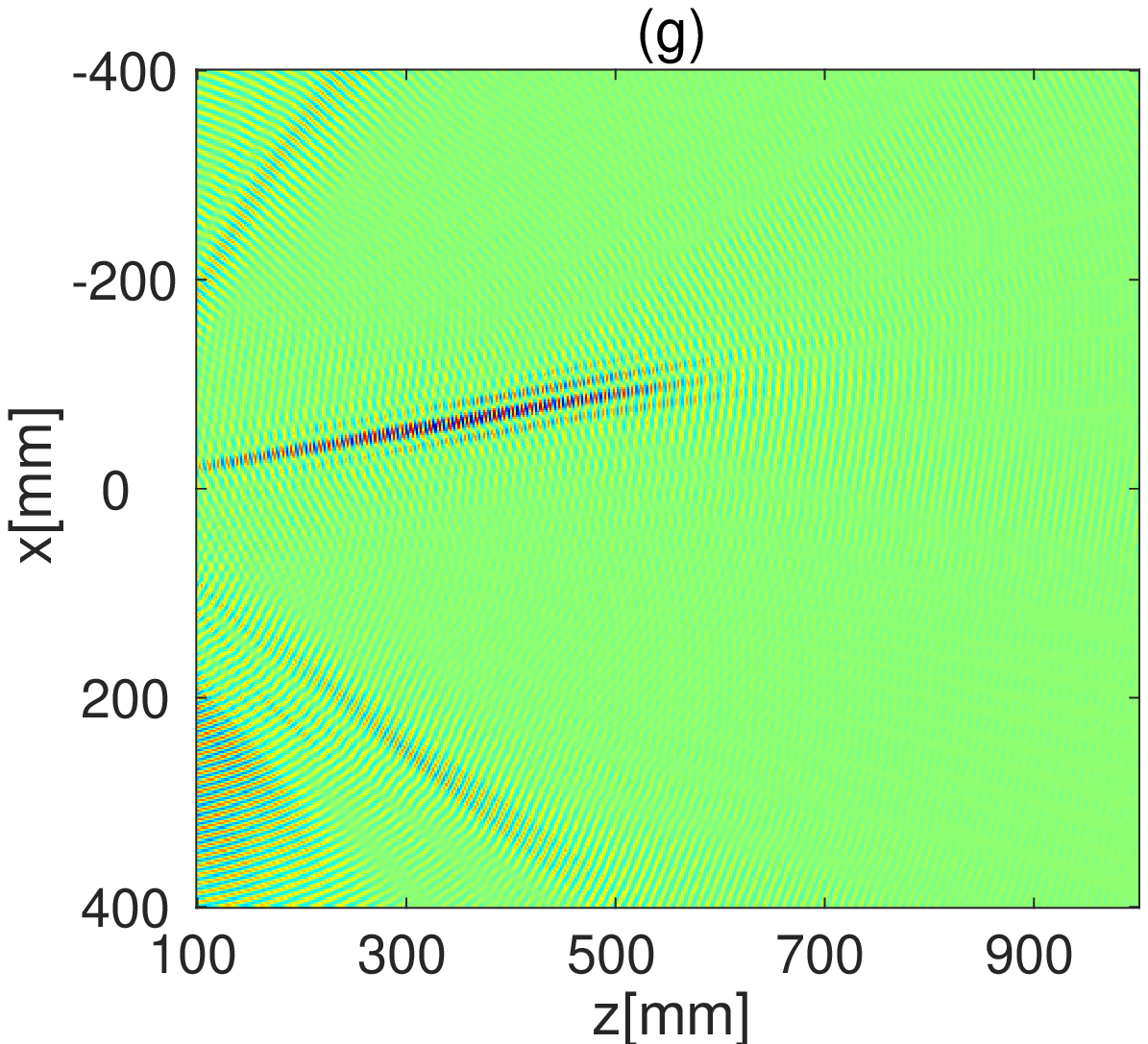}
\includegraphics[width=33mm]{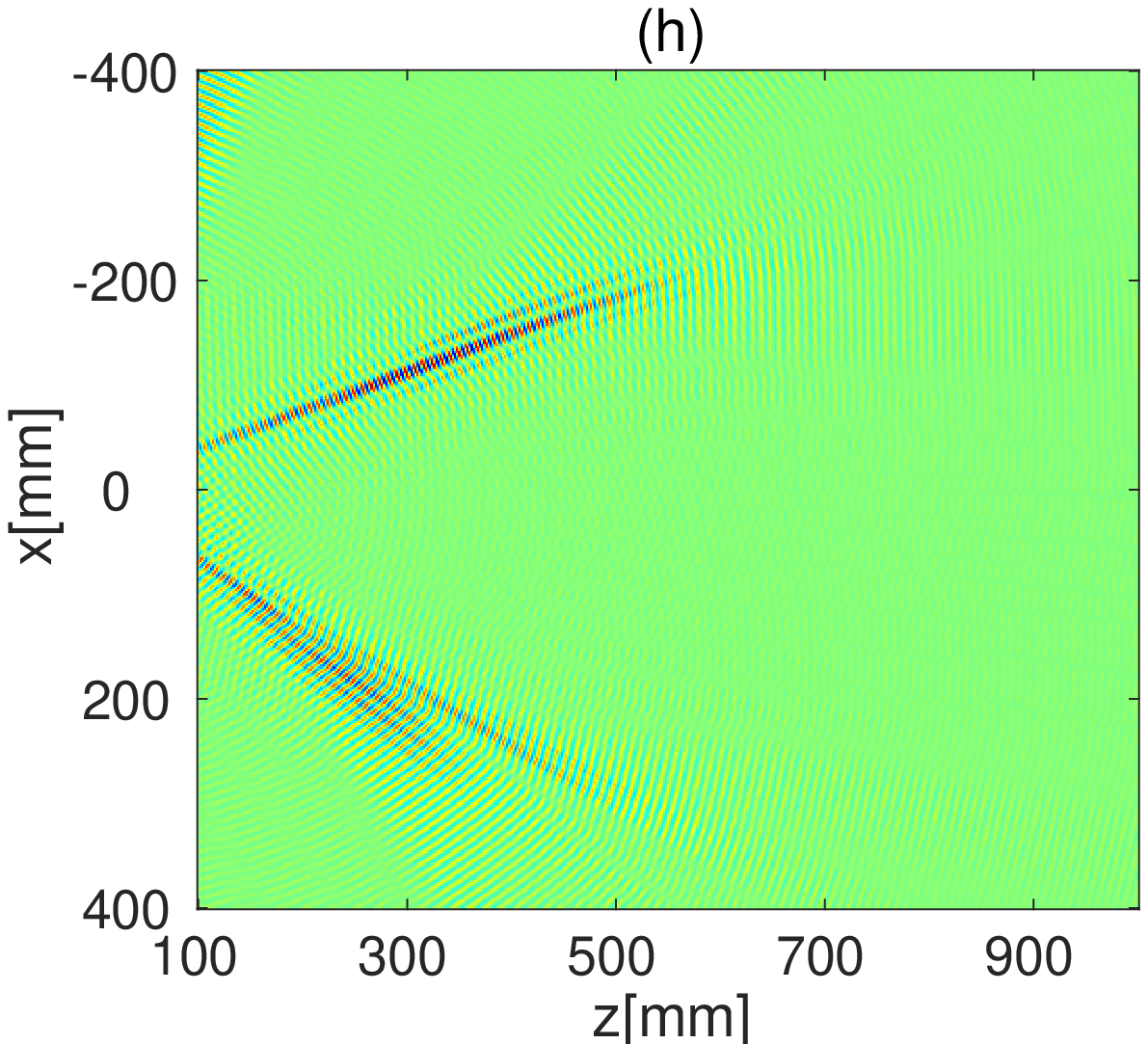}
\includegraphics[width=33mm]{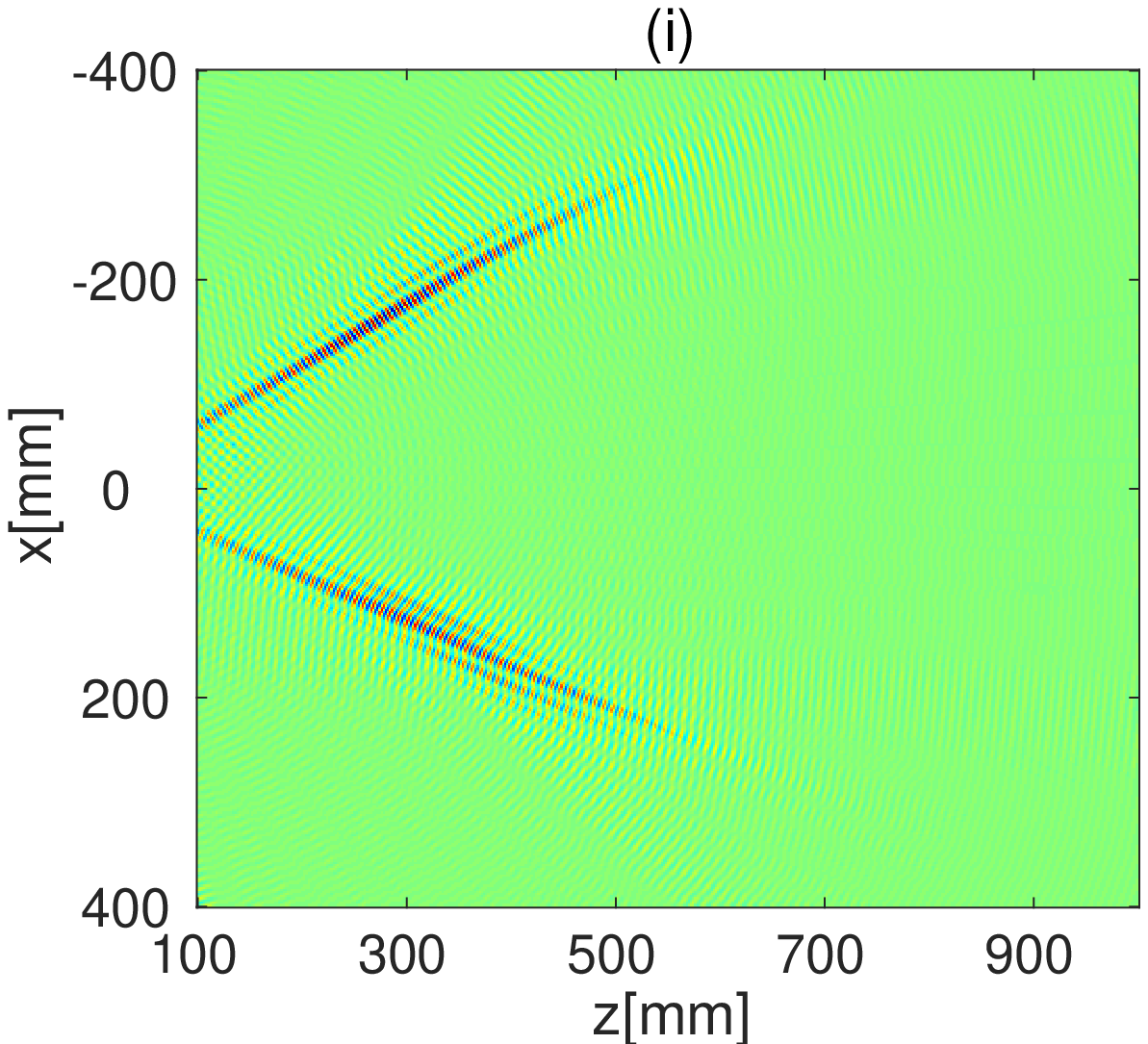}
\includegraphics[width=33mm]{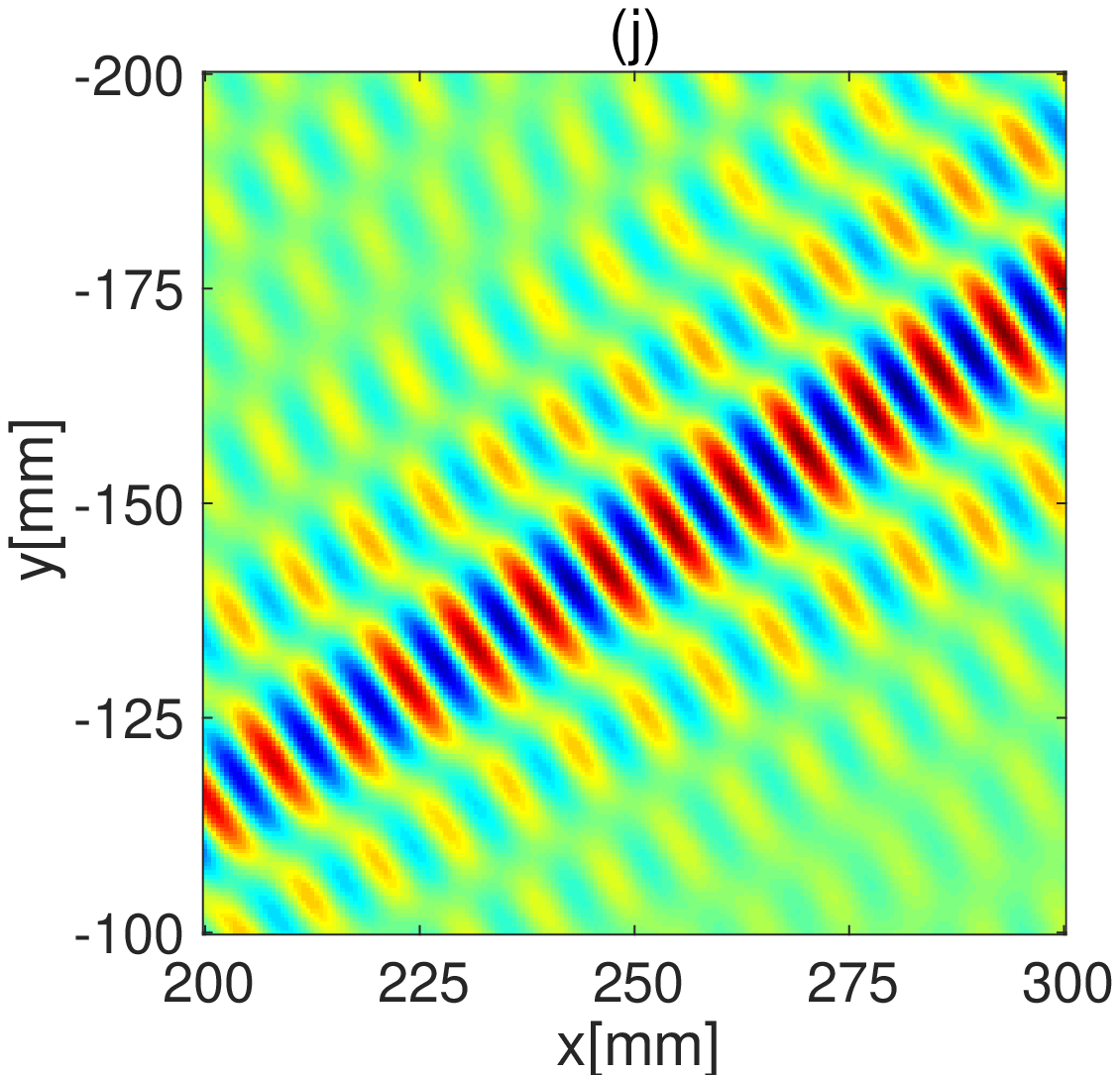}

\caption{(a)-(d): Fog guided along with beam propagation with tilting beam angle of (a)$0^\circ, (b)10^\circ, (c)20^\circ, (d)30^\circ$ from left to right. (e): Fog spreading without acoustic Bessel beams. (f)-(i): Numerical simulations showing acoustic pressure distribution corresponding to the case (a)-(d), respectively. (j): Magnified graph of (i) indicating the wavefront perpendicular to the beam orientation.}

\end{figure*}

The generated beam is thus demonstrated to be steerable, although its behavior in its 'fringe' would require further investigation.
Our technique only expresses the flow in the midst of the generated beam, not on the boundary of the beam or outside of it.
Since our measurement conducted in the paper does not contain the 'vector' information of the flow field, how particles in the periphery of the beam move remains unclear.
While taking pictures in Fig.4, we experienced that the fog was guided to the intended direction only when it is captured at the very center of the beam.
A slight deviation from the center would lead the fog toward the more or less tilted direction.
In fig.4(j), a pair of weaker beams parallel to the primary beam is seen, which is the characteristic energy distribution of Bessel beams.
Perhaps these secondary beams might cause the changes in traveling orientation when capturing the flow.

As we described above, there is a trade-off between the concentration and the reaching distance of the flow.
If one wants a well-concentrate flow over a long distance, operating a larger number of array will be one solution.
It is equivalent to prepare an acoustic lens with a greater aperture.
The upper limit for the beam concentration is given by its wavelength, 8.5~mm for the phased array used in the experiment.
A narrower flow would be possible with the use of ultrasound transducer with a higher output frequency.

In conclusion, we have described a method for generating ultrasonic Bessel beam whose propagation direction is steerable.
We have experimentally demonstrated that the sound beam generated with our method created a straight stream along with the ultrasound propagation.
The generated beam has its highest velocity apart from the sound source, which corroborated the mid-air acceleration of air particles inside the beam.
The flow was observed immediately after the beam is emitted.

The physical properties of the generated flows are to be investigated in greater detail.
For example, the vector information of the flow field would be the next step.
We are also curious if multiple flows with our technique can combine with each other and construct a complicated path of the airborne substances.

Since there have been techniques for generating other intriguing forms of sound propagations such as self-bending beams\cite{selfbend} and tractor beams\cite{tractor}, there will be a chance to generate even more complicated flows with our concept.

This work is supported from JSPS Kakenhi Grant-in-Aid for Young Scientists (A), 15H05316.

% Produces the bibliography via BibTeX.

\end{document}